\documentclass[12pt,twoside,letterpaper,twocolumn]{article}
\pdfoutput=1
\usepackage[margin=20mm]{geometry}

\usepackage{url}
\usepackage{graphicx}
\usepackage{amssymb,amsmath}

\begin{document}
\author{Ross Anderson and Robert Brady\\
        \small University of Cambridge Computer Laboratory\\
        \small JJ Thomson Avenue, Cambridge CB3 0FD, United Kingdom\\
        \small \texttt{\{ross.anderson,robert.brady\}@cl.cam.ac.uk}}
\title{Why quantum computing is hard\\-- and quantum cryptography is not provably secure}
\date{\today}

\twocolumn[
  \begin{@twocolumnfalse}
    \maketitle
\begin{abstract}
  Despite high hopes for quantum computation in the 1990s, progress in the past
  decade has been slow; we still cannot perform computation with more than
  about three qubits and are no closer to solving problems of real interest
  than a decade ago. Separately, recent experiments in fluid mechanics have
  demonstrated the emergence of a full range of quantum phenomena from
  completely classical motion. We present two specific hypotheses. First,
  Kuramoto theory may give a basis for geometrical thinking about entanglement.
  Second, we consider a recent soliton model of the electron, in which the
  quantum-mechanical wave function is a phase modulation of a carrier wave.
  Both models are consistent with one another and with observation.  Both
  models suggest how entanglement and decoherence may be related to device
  geometry. Both models predict that it will be difficult to maintain phase
  coherence of more than three qubits in the plane, or four qubits in a
  three-dimensional structure. The soliton model also shows that the
  experimental work which appeared to demonstrate a violation of Bell's
  inequalities might not actually do so; regardless of whether it is a correct
  description of the world, it exposes a flaw in the logic of the Bell tests.
  Thus the case for the security of EPR-based quantum cryptography has just not
  been made. We propose experiments in quantum computation to test this.
  Finally, we examine two possible interpretations of such soliton models: one
  is consistent with the transactional interpretation of quantum mechanics,
  while the other is an entirely classical model in which we do not have to
  abandon the idea of a single world where action is local and causal.
\end{abstract}
  \end{@twocolumnfalse}
  ]


\section{Introduction}
\label{sec:introduction}

Quantum computation appears straightforward at small scales of two or three
qubits, but attempts to scale it up have not been successful. Shor showed in
1994 that large-scale quantum computers could have significant impact, such as
in factoring the large integers that form the basis of RSA
cryptography~\cite{Shor}, but this would require maintaining coherence among
thousands of qubits. In 1998, Jones, Mosca and Hansen reported a quantum
computer with two qubits~\cite{JMH98} while Chuang, Gershenfeld, Kubinec and
Leung demonstrated a cascade of three~\cite{CGKL98}. In 2001, Vanderspysen,
Steffen, Breyta, Yannoni, Sherwood and Chuang reported a quantum computer that
could factor 15~\cite{Vanderspeisen+2001}. In 2002, the Los Alamos quantum
information science and technology roadmap aimed at having functioning quantum
computation testbeds by 2012~\cite{LAR}. See Chen et al~\cite{Chen+} for an
extensive survey of the technology.  Yet despite the investment of tremendous
funding resources worldwide, we don't have working testbeds; we're still stuck
at factoring 15 using a three-qubit algorithm~\cite{Lucero+2012}.

It is time to wonder whether there might be something we missed, such as
theoretical limits on entanglement and coherence.  Doubts about the feasibility
of quantum computers go back to 1995, when Unruh warned that maintaining
coherence might be hard~\cite{Unruh95}; researchers in this field still see the
problem in terms of reducing sources of noise (for example by using lower
temperatures), on increasing the signal (for example by bringing the particles
closer together) and on using error-correcting
codes~\cite{Chen+,Zurek2003}. Researchers are now starting to wonder whether
geometry affects entanglement and coherence; the first workshop on this topic
was held last year~\cite{GQE12}. However, experiments elsewhere in physics
suggest a type of limit that has not so far been considered.

\section{Guiding waves}
\label{sec:possible-limit-qubits}
 
\begin{figure}[h]
	\includegraphics[width=0.49\textwidth]{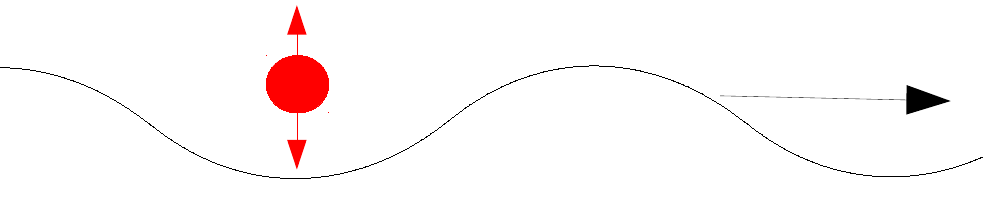}
	\caption{Schematic of droplet phase locked with surface waves}
	\label{fig:DropletPhaseLockedWithSurfaceWaves}
\end{figure}

In recent experiments by Couder and colleagues~\cite{Couder-nature,
  Couder-2006a, Couder-2006b, Couder-2009, Couder-2010}, a small liquid drop is
kept bouncing on the surface of a bath of the same liquid by oscillating this
substrate vertically.  The bouncing induces waves in the surface which, in
certain regimes, guide the motion of the droplet. As shown schematically in
Figure \ref{fig:DropletPhaseLockedWithSurfaceWaves}, in this regime the droplet
moves along the surface at the same velocity as the peaks and troughs of the
waves in the vicinity.

By measuring the statistical motion of the droplet, the experiments show clear
phenomena corresponding to those of quantum mechanics, including single-slit
diffraction, double-slit diffraction, quantised energy levels and tunnelling
through a barrier. A video shows clearly how quantum-mechanical phenomena can
arise in a completely classical system~\cite{Couder2011}.

In this two-dimensional analogue there is a limit to the number of qubits in a
coherent system. It is easy to get phase coherence with waves associated with
one other particle and possible to get coherence with two -- one coherence per
dimension. (In a three-dimensional system, a further coherence could be added.)
Kuramoto and others have developed extensive mathematical models of coupled
oscillators; for a review, see Acebr\'on et al.~\cite{Acebron+2007}. It is the
Dangelmayr-Knobloch radial standing-wave solutions that appear of most interest
here~\cite{DK87}. Even so, a single coherence between an ensemble of particles
is more likely, so that they will act as a single ensemble, as when the many
electrons in a Josephson junction act as a single qubit. (Coupled-oscillator
models have already helped explain other aspects of Josephson junction
behaviour.)

Couder's experimental measurements are also evocative of the de Broglie--Bohm
model of quantum mechanics~\cite{Albert, Bohm, Bell}, which is equivalent to
the traditional Copenhagen interpretation. In this model, a small particle
interacts with waves in three dimensions which obey the same equations as the
quantum mechanical wavefunction. The motion of the particle is given by
\begin{equation}
v ~= ~\frac{\hbar}{m} ~Im \left( \frac{\nabla \psi}{\psi} \right)
\label{eq:dbb}
\end{equation}   
and the resulting observables are the same as those of the Copenhagen
interpretation; in fact equation (\ref{eq:dbb}) is merely the equation that is
required for this to happen (it is derived from the usual quantum mechanical
wavefunction plus a continuity condition). The models are also equivalent for a
quantum mechanical system with entangled states.  Indeed Nikoli\'c has argued
that had the Bohm interpretation come along first, no-one would have needed the
Copenhagen interpretation~\cite{Nikolic2007}. But the de Broglie--Bohm model
may give more insight into what happens when a system loses coherence.

If two particles are entangled, then the guiding wave $\psi$ of one particle
must be correlated with that of the other. Now as quantum wavefunctions are
considered to be nonlocal, this caused difficulty for some writers: Bell, for
example, argued that the nonlocal nature of the wavefunction of two spin-$1/2$
entangled particles meant that a geometrical interpretation of the guiding wave
was impossible~\cite{Bell}. The textbook approach is that in such circumstances
the guiding wave is in six-dimensional configuration space, for which a
geometric interpretation in physical space is not obvious. Yet Bell also warned
that impossibility proofs mostly represented a failure of imagination, and he
himself had demolished previous arguments against a local-realist
interpretation of quantum mechanics.

We will argue, first, that the loss of phase coherence may provide a better
model for the behaviour observed in quantum decoherence experiments; and
second, that this hypothesis might be tested by decoherence experiments that
measure the physical geometry associated with entanglement and decoherence.
Before that, we will discuss how soliton models might provide some insight into
possible underlying mechanisms, in order to tackle the imagination failure. By
presenting a local-realist model that is consistent with de Broglie--Bohm and
with observed empirical results, we challenge the argument of impossibility.

\section{Soliton models}

Solitons are persistent, localised solutions of the wave equation (with
additional nonlinear terms, which are usually small). They arise in fluid and
other media, having first been observed and described on a canal in the
mid-19th century~\cite{solitons}, and were applied to particle physics
following the proposal by Skyrme in 1961 of a model of an atomic nucleus, later
developed and popularised by Witten~\cite{Skyrme,Witten}. Many other soliton
models have been proposed in various branches of physics. More recently, for
example, Volovik has found that quasiparticles in liquid helium exhibit many of
the properties described by the Copenhagen model and relativity (albeit with
$c$ being the speed of sound in the fluid)~\cite{Volovik}, and raised the
question of whether fluid models could be applied to all elementary particles.

In the field of analogue gravity, Unruh and others have explored fluid models
of black holes~\cite{Unruh81} and this led to a thriving research programme
exploring many provocative analogies between fluid flow and general
relativity~\cite{Barcel11}. In particular, an event horizon corresponds to the
start of supersonic flow; Lahav and colleagues have observed this
experimentally in a Bose-Einstein condensate~\cite{Lahav+2010}. In short, over
the past thirty years, fluid models have developed to express most of the
properties of elementary particles from the basic Copenhagen model to (in
aggregate) general relativity.

In a companion paper, Brady has proposed a soliton model for the
electron~\cite{Brady} which we will now summarise. It provides a fluid-model
analogue of the Coulomb force, and is thus of relevance at least to decoherence
in quantum computers relying on electron behaviour (such as qubits based on
Josephson junctions).  The key insight is that Euler's equation for a
compressible fluid possesses quasiparticle solutions with chirality. These may
be visualised as smoke rings but with a twist, in that the line of greatest
pressure circulates not merely around the ring's long diameter but around its
short one too.

Consider a compressible inviscid fluid of pressure $P$, density
$\rho$ and velocity ${\bf u}$ of an inviscid fluid medium that obeys Euler's
equation:
\begin{equation}
	\frac{\partial {\bf u}}{\partial t} + ({\bf u}. \nabla){\bf u} = - \frac{1}{\rho}\nabla P 
	\label{eq:euler}
\end{equation}
where $\partial \rho / \partial t = -\nabla (\rho {\bf u})$. At low
amplitude, this gives the wave equation
\begin{equation}
	\frac{\partial^2 \rho}{\partial t^2} = c^2\nabla^2 \rho
	\label{eq:wave}
\end{equation}

The wave equation has linear solutions, and also eddy-like solutions like smoke
rings. There the line of greatest density rotates round the ring's small axis,
as in Figure 2a. However, there are also chiral solutions where the line of
greatest density rotates around both axes, as in figure 2b. The general
solutions are referred to as sonons. This solution of the wave equation can be
written
\begin{equation}
	\xi_{mn} = \psi_o R_{mn}
	\label{sonon}
\end{equation}
where
\begin{equation}
	\psi_o = Ae^{-i\omega_0t}
	\label{eq:psi-o}
\end{equation}
\begin{equation}
R_{mn} = \int_0^{2 \pi} e^{-i (m \theta' - n \phi)} j_m (k_r \sigma) k_r R_o d\phi
\label{eq:rmn}
\end{equation}
\begin{figure}[htb]
	\centering
		\includegraphics[width=1.00\columnwidth]{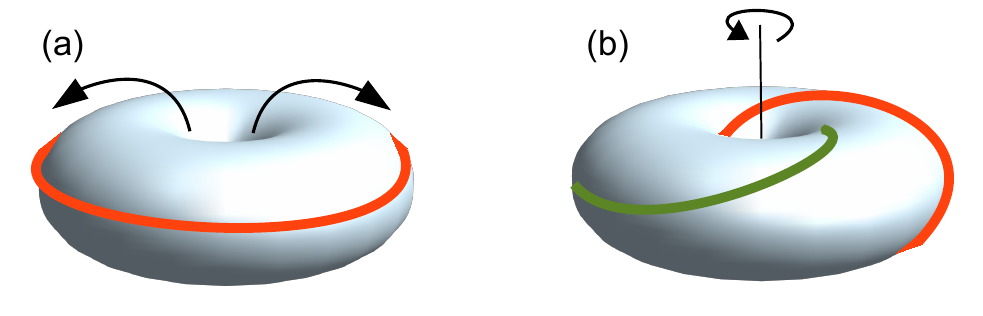}
	\caption{Sonons (a) without chirality (b) with chirality}
	\label{fig:two-sonons}
\end{figure}

Figure \ref{fig:two-sonons}a shows the $R_{10}$ sonon. The red line is the line
of maximum density, rotating at angular speed $\omega_0$. Figure
\ref{fig:two-sonons}b shows the $R_{11}$ sonon, which models the electron. In
such particles, the chirality, spin direction, $m$ and $n$ are preserved by
continuous transformations, so are persistent and quantised. At low amplitude
they are Lorentz covariant because they obey the wave equation \eqref{eq:wave},
which is Lorentz itself covariant, and it turns out that the perturbations at
finite amplitude average to zero over a cycle. Classical dynamics follow in the
approximation of constant $R_{mn}$ and small $v/c$. Meanwhile, at a large
distance from the sonon, $\chi$ may be approximated up to a phase factor as
\begin{equation}
	\chi = \frac{1}{r} \sin k_r r
	\label{eq:chi-large-r}
\end{equation}
(We refer the reader to~\cite{Brady} for the details.) 

The important point for this paper is that $\chi$ behaves like a carrier wave
and $\psi$ as its modulation, which is a complex function as its phase is
important. This provides a physical model of the de Broglie--Bohm view that a
particle moves through space surrounded by waves that obey the usual quantum
equations. Extending equation \eqref{eq:psi-o} into a Lorentz covariant form leads directly to the Klein--Gordon equation
\begin{equation}
  \frac{\partial^2 \psi}{\partial t^2} -c^2 \nabla^2 \psi = - \omega_0^2 \psi 
\end{equation}
(the relativistic form of Schr\"odinger's equation); with a little more work we
find that the $R_{11}$ sonon is governed by the Dirac equation, which describes
the behaviour of the electron in detail~\cite{Brady}. It follows that provided
a system remains coherent, the usual predictions of quantum mechanics will
apply.  (The analogue gravity community has found numerous cases of quantized
behaviour of sound waves in fluids and applied them as analogies to other
problems in quantum physics; see the survey by Barcel\'o, Liberati and
Visser~\cite{Barcel11}.)

The more detailed equations (4--7) enable us to make a number of predictions
about decoherence. For example, as the carrier wave $\chi$ decays as $1/r$,
the system will be more prone to decoherence with distance. 


In the absence of decoherence, the equations of motion are time-reversal
symmetric, as Euler's equation is. The state of the system at any one time
determines its state at any other time, whether in the future or in the
past. Thus it might not be surprising if we see behaviour that appears to
violate microcausality~\cite{Bennett87}; entropy kicks in once phase coherence
is lost.  The big question is whether we can have a local realist model of
quantum systems without violation of macrocausality. This leads us to Bell's
theorem.

\section{Local realism and quantum crypography}

If the soliton model of the electron (or perhaps another coupled-oscillator
theory) is correct, then two of the possibilities are as follows.

\begin{description}
\item [Weak (transactional) soliton hypothesis:] the elementary particles
  are solitons in an inviscid fluid, but time reversal symmetry in entangled
  states means that there may be violations of microcausality. We still get
  quantum electrodynamics with advanced and retarded waves following the
  exposition of Mead~\cite{Mead2000}, and relativity works because all
  particles are solutions to the wave equation and thus Lorentz covariant.
\item [Strong (causal) soliton hypothesis:] the elementary particles are
  solitons in an inviscid fluid; relativity emerges from the fact they satisfy
  the wave equation; and quantum mechanics from the nature of the solutions. So
  Euler's equation explains not just the motion of matter, but also
  electricity, light and atomic forces.
\end{description}

These two interpretations give quite different views of reality. The first is
analogous to Cramer's transactional interpretation of quantum
mechanics~\cite{Cramer86}. The second is a classical view of the world;
Newton's laws determine everything, including the very large and the very
small.

Initially one might think that Bell's theorem, and the entanglement experiments
inspired by it, compel us to favour the former. But a closer examination
suggests that this is not necessarily so, because the experiments are designed to interact with the propagating waves, not, on this hypothesis, with the carrier waves which might themselves carry information about spin correlations.

If an experimenter creates a pair of entangled particles, sends one of them
round an optical fibre or waveguide or tunnel of length $D$, and then performs
a measurement on them with equipment spaced a distance $d$ apart for the two
particles, then although the $\psi$ waves of the soliton may have travelled a
spacelike separation $D$, this does not necessarily hold for the $\chi$ waves
whose phase coherence creates the entanglement in the soliton model. The $\chi$
waves are broadcast in all directions from a sonon and thus the distance that
matters to prove impossibility results about coherence is $d$.  If this is not
spacelike then no violation of locality (or relativity or causality) has been
proved.

In 1982, Aspect, Dalibard and Roger tried to exhibit a spacelike separation by
using polarisers that switched in 10ns while the length L of the path traversed
by the photons had $L/c$ = 40ns~\cite{Aspect1982b}. Yet they used a single
receiver for coincidence monitoring, so $d$ = 0. 

In 1998, Tittel, Brendel, Zbinden and Gisin demonstrated coherence in photons
sent round a 10.9km optical fibre in a direct attempt to probe the tension
between quantum non locality and relativity; yet the same issue arises with
this experiment~\cite{Tittel+98}. The source, located in Geneva, was 4.5 km
from the first analyser in Bellevue and 7.3 km from the second in Bernex, with
connecting fibers of 8.1 and 9.3 km. However, entangled states were studied
only when both photons went either through the short arms or through the long
arms.

In the same year, Weihs, Jennewein, Simon, Weinfurter and Zeilinger performed
an experiment with what they believed was a proper spacelike separation: photon
pairs were sent from a source to two detectors 400m apart and were found to be
coherent on arrival~\cite{Weihs+98}. However this does not establish that
information was transmitted faster than light by the $\psi$ wavefuntion, as
coherence is maintained by the $\chi$ wave which travels at the speed of light
just like the photons but in a straight line. 

In 2008 Salart, Baas, van Houwelingen, Gisin and Zbinden did a fibre-loop
experiment over a distance of 18km (from Geneva to Satigny and Jussy) and
actuated a piezoelectric crystal which moved a mirror, ensuring that coherence
was lost~\cite{Salart+2008}; yet the same applies here as in Weihs' experiment.

In short, experimenters have sought to close one loophole after another in the
Bell test experiments over the last thirty years. But the soliton model of the
electron creates another major hole as the experimenter must consider not just
the propagation of the quantum-mechanical wavefuntion $\psi$ but also of the
density waves $\chi$ on which they are modulated.

The consequences for quantum crypto are notable. As the experiments done to
test the Bell inequalities have failed to rule out a classical hidden-variable
theory of quantum mechanics such as the soliton model, the security case for
quantum cryptography based on EPR pairs has not been made.

We propose that experimeters test explicitly whether entanglement is a function
of physical geometry in the way predicted by the soliton model, or more
generally by the results of Kuramoto theory.

First, one might fabricate a series of 3-qubit quantum computers with the
coherent elements in a triangle whose largest angle was 90$^o$, 100$^o$, ...,
180$^o$. We predict that 3 distinct qubits will not be measured when the
elements are collinear, and perhaps also when they are nearly collinear. One
might also make a 4-qubit machine in three dimensions, and similarly measure
the correlation with geometry.

Second, more general entanglement experiments might attempt to identify
behaviour consistent with Kuramoto theory such as finite size effects on
decoherence, relationships with the order parameter and whether bifurcation
points can explain the circumstances in which systems become coherent.

Third, we suggest close scrutiny of claims that computation can be sustained
without decoherence. If the strong soliton hypothesis is correct, we would
expect that a single physical qubit cannot be recycled in the same coherent
computation; thus if a computation requires $k$ steps on $n$ qubits it would
need at least a $k$-by-$n$ array of qubits, not a single $k$-qubit register
plus some CNOT gates.  If quantum mechanics is really just a convenient
calculus for dealing with coupled oscillators, then reality is classical, and
quantum computers are just classical computers. They cannot then provide a way
to beat the Bremermann limit of $mc^2/h$ computations per second for a computer
of mass $m$~\cite{Bremer65}.

\section{Conclusions}
\label{sec:conclusion}

One of the big puzzles that straddles the boundary between physics and
computation is why quantum computers have got stuck at three qubits. We have
shown that a local-realist version of the de Broglie--Bohm interpretation of
quantum mechanics provides a good explanation: entangled particles are
precisely those whose guiding waves are phase coherent. It follows that we can
expect two entangled qubits to be possible on a line, three in a plane and four
in a three-dimensional structure. In fact, it may be more helpful to model
qubits as coupled oscillators, following Mead's model of quantum
electrodynamics and Kuramoto theory, than using Hilbert space. We propose
experiments to verify this directly.

Bell warned that claimed impossibility proofs often showed merely a lack of
imagination on the part of the `prover', so we presented a concrete
guiding-wave model given by a recent soliton model of the electron. In this
model, the electron is a spinning twisted torus in an inviscid fluid. It
generates compression waves $\chi$ which are in turn modulated by guiding waves
$\psi$.

Since the Bell test community has not yet considered the possibility that
coherence information might be transmitted other than by the quantum mechanical
wavefunction $\psi$, the experiments that have claimed to demonstrate nonlocal
behaviour of entangled systems have done nothing of the kind.  If entanglement
is simply phase coherence, it is not enough to show that two photons sent to
separated sensors remain coherent even though the distance between the sensors
have a spacelike separation, as the phase coherence is carried by the $\chi$
waves.  In consequence we dispute the claim that a quantum cryptosystem based
on EPR pairs must be secure. The evidence needed to support that has simply
never been exhibited.

We also challenge experimentalists who believe that entangled states violate
locality to devise an experiment where locality fails in the soliton model. In
fact since quantum mechanics and relativity can both be derived from this local
and causal model, it will be surprising if anyone can use Bell's theorem to
prove an incompatibility betweeen quantum mechanics, relativity, locality and
causality, regardless of whether the soliton model turns out in the end to be
the right one.  More generally, we invite experimentalists to investigate the
physical geometry of entanglement and coherence. The real prize is not the
ability to build better quantum machines, but the far greater one of
understanding the most fundamental questions. Do soliton models provide a
better explanation of the world than string theories?  If so, which soliton
models are supported?  And in the absence of evidence, we need not accept that
physics really requires us to abandon the concept of a single objective
universe where action is both local and causal.

\small

\end{document}